\documentclass[twocolumn,pra,floatfix]{revtex4}

\usepackage{physics}
\usepackage{color}
\usepackage{tabularx}
\usepackage{epsfig}
\usepackage{amsmath}
\usepackage{amssymb}
\usepackage{bm}
\usepackage{graphicx}
\usepackage{multirow}

\usepackage{epsfig}

\begin{document}

\title{The Role of Geometric Analysis in Interferometer Design and Optimization for Gravitational Quantum Entanglement}

\author{Alireza Maleki}
\affiliation {School of Physics, Institute for Research in Fundamental Sciences (IPM), Tehran, Iran} 

\begin{abstract}
The pursuit of a quantum theory of gravity, aiming to unify general relativity and quantum mechanics, remains one of the most enduring challenges in physics. Because of the extreme energy scales associated with the Planck regime, direct experimental evidence for quantum gravity remains elusive. However, recent proposals suggest that quantum entanglement between two massive particles may provide a pathway to probe the quantum nature of gravity. In this study, we examine the interferometer geometries proposed in these works, with particular attention to the commonly used approximation that neglects phase contributions from the vertical segments of the particle trajectories. Our analysis shows that this approximation can lead to incorrect predictions and, in certain parameter regimes, to null results where entanglement would otherwise be expected. We derive exact solutions that incorporate the full particle trajectories and demonstrate that the vertical arms can significantly affect the accumulated phase. Crucially, we identify configurations in which the induced entanglement vanishes entirely, a feature missed by simplified treatments. These findings show that accounting for the full interferometer geometry is not merely a refinement, but is essential for accurately assessing gravity-induced entanglement.	
\end{abstract}

\maketitle

\section{Introduction}
Modern physics is built upon two remarkably successful theories of quantum mechanics and general relativity, jointly forming the foundation of our understanding of phenomena ranging from the microscopic realm of particles to the vast scales of cosmic structures \cite{lawrie2012unified,buchbinder2021introduction,maleki2020speed,maleki2024cosmic,maleki2026area}.
Yet, they are fundamentally incompatible. At the Planck scale, general relativity breaks down, and quantum theory provides no description of spacetime. 
Singularities and paradoxes, such as the black hole information problem—rooted in quantum effects near black hole horizons and their eventual evaporation underscore the pressing need for a quantum theory of gravity \cite{hawking1974black, marletto2025quantum}. Over the years, some theoretical frameworks, including string theory and loop quantum gravity, have been proposed to address this challenge \cite{kiefer2007quantum,rovelli2008loop}. However, experimental tests remain extremely difficult because Planck-scale energies are far beyond our reach  \cite{alfaro2005quantum}. Some approaches suggest that gravity may be an emergent phenomenon rather than a fundamental interaction \cite{padmanabhan2015emergent,verlinde2017emergent}. These ideas make experimental tests all the more important for distinguishing between classical and quantum descriptions of gravity. As one of the pioneering efforts in this direction, Richard Feynman’s 1957 proposal for a Stern–Gerlach-type experiment involving quantum spin superpositions explored the possibility of coupling gravitational interactions to quantum phenomena, emphasizing the need to quantize gravity \cite{cecile2011role}.

The significant technological progress of quantum information  science in recent years has inspired experimental strategies to probe the quantum nature of gravity \cite{peres2004quantum,lunghi2013experimental}. Among these are proposals that utilize phase shifts induced by gravitational interactions in setups like Stern-Gerlach and Mach-Zehnder interferometers to explore gravity's quantum characteristics \cite{bose2017spin,marletto2017gravitationally}. Specifically, these proposals aim to detect gravitationally induced quantum entanglement between two massive particles \cite{bose2017spin,marletto2017gravitationally,lin2026can}. The central premise is that the observation of such entanglement serves as evidence of gravity's quantum nature \cite{bose2025massive, marletto2025classical}.
These ideas have ignited optimism for the experimental detection of quantum gravitational effects \cite{carney2021using,krisnanda2020observable,pedernales2022enhancing,maleki2022complementarity}. However, despite their promise, these experimental proposals face significant theoretical and practical challenges.

The experimental realization of gravitationally induced entanglement typically requires massive quantum systems in the approximate range of $10^{-14}$--$10^{-12}\,\mathrm{kg}$ prepared in spatial superposition states at relatively short separations, the regime where gravitationally generated phases become appreciable \cite{bose2017spin,marletto2017gravitationally,schut2023relaxation,weber2024bose,di2024bose,bose2509spin}. The desired signal competes with several sources of decoherence and noise, including scattering with molecules of air, thermal photons, thermal vibrations, non-uniformities in the gravitational field, and perturbations due to nearby massive objects \cite{gunnink2023gravitational,marletto2025quantum}. Electromagnetic interactions such as Casimir--Polder forces at short separations, electrostatic potentials from residual charging, and magnetic dipole interactions for particles with spin can dominate over gravity or introduce phase shifts that may be difficult to distinguish from the gravitational contribution \cite{schut2023relaxation,bose2025massive}. These effects impose stringent experimental requirements ultra-high vacuum, cryogenic temperatures, electromagnetic shielding, and advanced vibration isolation and constrain the feasible parameter space for generating detectable entanglement \cite{bose2025massive,schut2023relaxation,marletto2025quantum}.

Despite these challenges, the recent progress in Stern--Gerlach interferometry, spatial superposition techniques, and matter-wave interferometry \cite{margalit2021realization, keil2021stern, pedalino2026probing, arndt2014testing, bild2023schrodinger, inarrea2025schrodinger} suggests that these requirements are no longer purely speculative thought experiments, but are progressively becoming experimentally structured programs. Matter-wave interferometry has been successfully demonstrated with complex molecules up to approximately $10^{-21}$ kg \cite{pedalino2023experimental}. Although these masses remain far below the required range, ongoing experimental progress in controlling increasingly large quantum systems represents an encouraging trajectory toward the ultimate goal.

In this work, we revisit proposals aimed at detecting gravitationally induced quantum entanglement and critically analyze the commonly adopted approximations. By developing exact solutions, we demonstrate how these approximations can substantially influence experimental outcomes. Furthermore, we explore the role of geometric configurations in the proposed experiments and identify optimized setups that significantly enhance their feasibility. Our findings not only refine the theoretical framework for these experiments but also provide practical strategies to strengthen their reliability and accuracy. Addressing these challenges is essential for advancing the experimental investigation of quantum gravity and realizing these experiments in laboratory settings.



 \section{The First Experimental Setup}  
 \label{Nstatecomplementarity}  

\begin{figure}[h]
 	\centering
 	\includegraphics[width=3.5in]{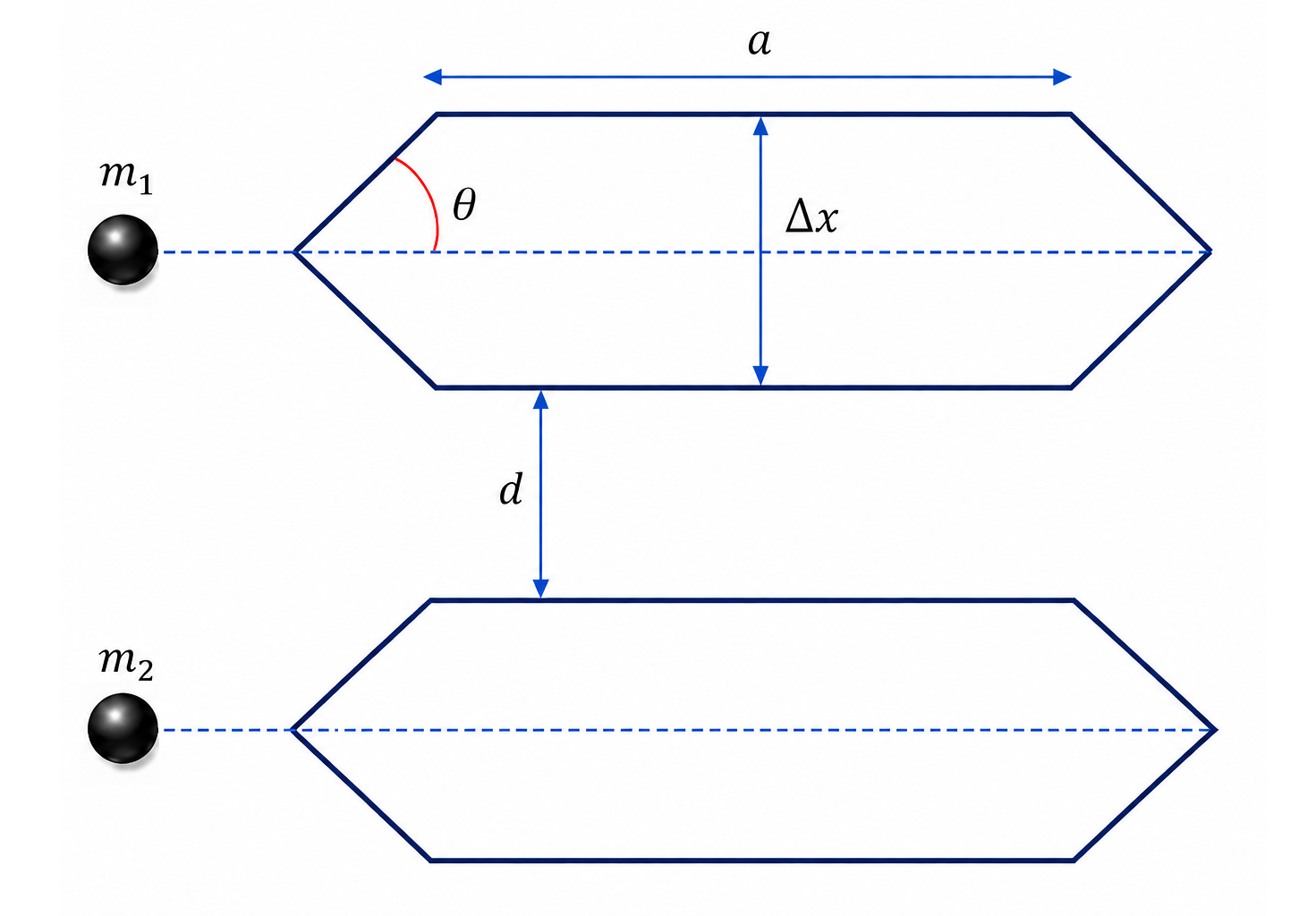}
 	\caption{Experimental setup for generating gravitationally induced quantum entanglement using two interferometers for massive particles in  Stern-Gerlach interferometers.}
 	\label{fig1}
 \end{figure}

 To analyze the quantum properties of gravity, we consider a setup involving two massive particles, each passing through a Stern-Gerlach interferometer, as shown in Fig. \ref{fig1}. This approach has been explored in recent studies \cite{bose2017spin,van2020quantum,margalit2021realization,schut2023relaxation,japha2023quantum, feng2024quantum}.
 
 In this experimental configuration, the two paths of the interferometer are separated by an angular difference of $2\theta$, with a vertical separation $\Delta x$. Throughout the remainder of the paper $\theta$, is referred to as the deviation angle. The parallel horizontal segment has a length of $a$, and the total path length traversed by a particle before detection is $l = a + \Delta x / \sin \theta$. Two identical Stern-Gerlach interferometers are assumed and placed at a distance $d$ from each other.
 
 The upper and lower paths in $i$-th interferometer ($i=1,2$), are associated with the orthogonal states $\vert u_i\rangle$ and $\vert d_i\rangle$. 
 
 To study gravitationally induced entanglement, we consider two massive particles, $m_1$ and $m_2$, each traversing their respective interferometers. At the initial path separation in each interferometer, the particles have probabilities of taking the upper or lower path. Considering that  the $i$-th interferometer has reflectivity and transmissivity parameters $(r_i, t_i)$. The initial state of the system, after passing through the first beam splitters, is given by

 \begin{equation}
 	\vert \psi(t=0)\rangle = \big(r_1\vert u_1\rangle + t_1\vert d_1\rangle\big) \otimes \big(r_2\vert u_2\rangle + t_2\vert d_2\rangle\big).  
 	\label{1ifnt}
 \end{equation}

Gravitational interactions have been shown to induce phase shifts in the quantum states of particles, as first demonstrated in the pioneering Colella-Overhauser-Werner (COW) experiment \cite{colella1975observation}. The gravitational phase shift is given by 

\begin{align}  
	\phi_{G} = -\int \frac{m \Phi}{\hbar} \, dt,  
	\label{Gphase}  
\end{align}  

where $m$ is the particle's mass, $\Phi$ is the gravitational potential affecting the particle, and $\hbar$ is the reduced Planck constant.

In our proposed setup, two particles, each confined to its own interferometer and separated by a fixed distance, experience gravitational interactions that induce phase shifts in their quantum states \cite{bose2017spin,marletto2017gravitationally}. This arrangement provides an opportunity to investigate quantum effects arising from gravitational interactions in detail.

Once the gravitationally induced phase shift is established, the gravitational interaction effectively decouples from the system \cite{bose2017spin,marletto2017gravitationally}. Consequently, the initial quantum state of the system, $|\psi(t=0)\rangle$, evolves into the state

\begin{align}
	\nonumber
	|\psi(t=T)\rangle &=  t_1t_2 e^{i\phi_1}|d_1\rangle |d_2\rangle + t_1r_2 e^{i\phi_2}|d_1\rangle |u_2\rangle \\ 
	&\quad +r_1t_2 e^{i\phi_3}|u_1\rangle |d_2\rangle + r_1r_2 e^{i\phi_4}|u_1\rangle |u_2\rangle,
	\label{3ifnt}
\end{align}

just before the path of the particles meets again at the end of the interferometer. 

In this configuration, it is commonly assumed that the interferometers are significantly longer than they are wide. Consequently, the primary contribution to the accumulated phase arises from the horizontal segments of the paths. The phases accumulated during the horizontal traversal are typically expressed as 

\begin{align}
	\nonumber
	\phi_1 &= \phi_4 = G\frac{m_1m_2T}{\hbar (d+\Delta x)}, \\ 	\nonumber
	\phi_2 &= G\frac{m_1m_2T}{\hbar d}, \\
	\phi_3 &= G\frac{m_1m_2T}{\hbar (d+2\Delta x)},
	\label{pHIapprox}
\end{align}

where $G$ is the gravitational constant, $T = l/v$ is the total time spent in the interferometer, $m_1$ and $m_2$ are the particle masses, $d$ is the separation between the interferometers, and $\Delta x$ is the horizontal path separation within each interferometer. 

In many analyses, the contributions from the vertical path segments are neglected. However, for small values of $\Delta x$, the paths may not separate sufficiently to induce a large enough phase difference to generate detectable entanglement. Conversely, for larger values of $\Delta x$, the contributions from the vertical paths may become significant and can no longer be ignored. This raises the question of the optimal configuration to maximize entanglement for a given total path length.

To address this, we consider an exact solution incorporating the contributions from the vertical paths. For particles traveling at a constant speed $v$ along their paths, the gravitationally induced phases are calculated as follows:

\begin{align}
	\phi_1 &= \phi_4 = G\frac{m_1m_2}{\hbar v} \left(\frac{\Delta x}{\sin \theta} + a \right) \frac{1}{d+\Delta x}, \\
	\phi_2 &= G\frac{m_1m_2}{\hbar v} \left( \frac{1}{\sin \theta} \ln \left(1+ \frac{\Delta x}{d} \right) + \frac{a}{d} \right), \\
	\phi_3 &= G\frac{m_1m_2}{\hbar v} \left( \frac{1}{\sin \theta} \ln \left(1 + \frac{\Delta x}{d+\Delta x} \right) + \frac{a}{d+2\Delta x} \right),
\end{align}

In the limit of small $\Delta x$ these expressions reduce to Eq.~\ref{pHIapprox}. The primary experimental challenge lies in achieving sufficiently large entanglement to be observable. To better understand this limitation, we analyze the entanglement in the system using a suitable entanglement measure.

The entanglement of a two-qubit quantum state $\rho$ can be quantified using the concurrence, which is defined in terms of the $R$ matrix as $R = \sqrt{\sqrt{\rho} \bar{\rho} \sqrt{\rho}}$, where $\bar{\rho} = (\sigma_y \otimes \sigma_y) \rho^\star (\sigma_y \otimes \sigma_y)$. The concurrence is calculated using the eigenvalues $\lambda_i$ of $R$, arranged in descending order, and is expressed as \cite{wootters1998entanglement}
\begin{align}
	\mathcal{C} = \text{max}\{0, \lambda_0 - \lambda_1 - \lambda_2 - \lambda_3\}.
\end{align}

From the evolved state, the concurrence simplifies to 
\begin{align}
	\mathcal{C} = 4(r_1r_2t_1t_2) \vert \sin(\Delta\phi/2)\vert,
\end{align}

where $\Delta\phi = 2\phi_1 - \phi_2 - \phi_3$. The concurrence achieves its maximum value when the beam splitters are configured with equal probability $r_i = t_i = 1/\sqrt{2}$, corresponding to 50:50 beam splitting. In this case, the concurrence reduces to $\mathcal{C} = \vert \sin(\Delta\phi/2)\vert$.  This value varies between 0 and 1, exhibiting an oscillatory behavior. The entanglement reaches a maximum for odd multiples of $\Delta\phi = k\pi$ (where $k$ is an odd integer) and vanishes for even multiples. To experimentally verify gravitationally induced entanglement, achieving maximum entanglement may not be necessary; a sufficiently large degree of entanglement can serve as evidence.

Experimental feasibility of gravitationally induced entanglement depends critically on the factor $\kappa = \frac{G m_1 m_2}{\hbar v}$. The key challenges involve achieving sufficiently large particle masses ($m_1, m_2$) and long interaction times ($T$) to accumulate significant phase differences. However, the role of the geometric configuration of the setup, as encapsulated in the phase combination, is often overlooked.

Optimizing the geometric configuration of the interferometers can substantially enhance the induced phase and, consequently, the entanglement. For this analysis, the total path length of the particles (the sum of the vertical and horizontal arms) is assumed to be fixed. To maximize entanglement, we optimize the vertical arm length $\Delta x$ and the deviation angle $\theta$. Since $\kappa = \frac{G m_1 m_2}{\hbar v}$ is common to all phases, the rescaled phase factor $\Delta\phi/\kappa$ is analyzed.
\begin{figure}[t]
	\centering
	\includegraphics[width=3in]{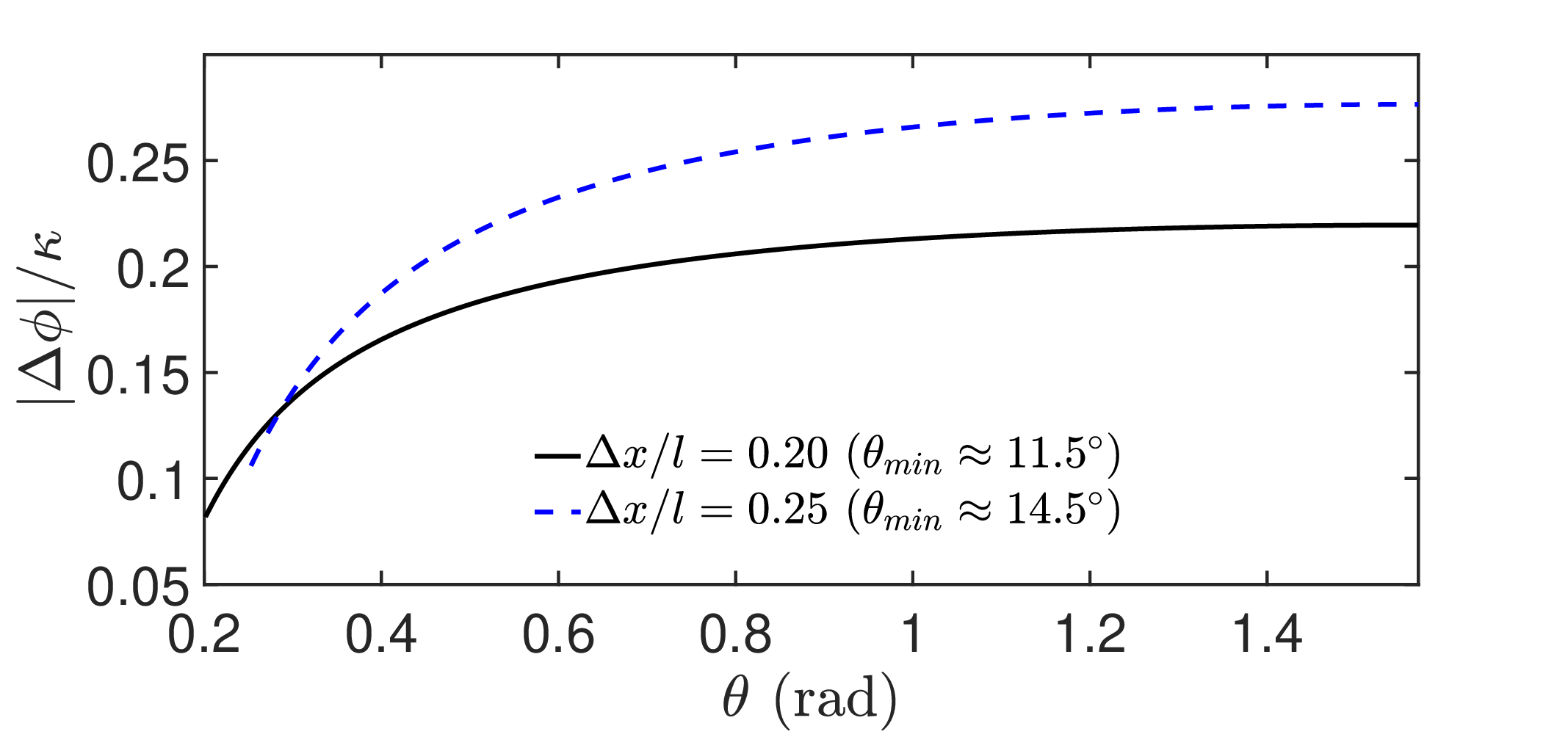}
	\caption{Rescaled induced phase $|\Delta\phi|/\kappa$ as a function of the deviation angle $\theta$ for $\Delta x/l = 0.20$ and $\Delta x/l = 0.25$ at fixed $d/l=0.5$. The allowed region is restricted by the geometric constraint leading to  $\theta_{min}= \arcsin(\Delta x/l)$.}
	\label{dxThetaBose}
\end{figure}

Figure~\ref{dxThetaBose} shows the dependence of the rescaled phase shift on the deviation angle $\theta$ for fixed $\Delta x/l$ and $d/l$. The lower bound on $\theta$ follows from the geometric relation $\Delta x/l=(1-a/l)\sin\theta$, which implies $\theta>\arcsin(\Delta x/l)$. As $\theta$ increases, the magnitude of the phase grows and approaches its maximum near $\theta=\pi/2$.

\begin{figure}[t]
	\centering
	\includegraphics[width=3in]{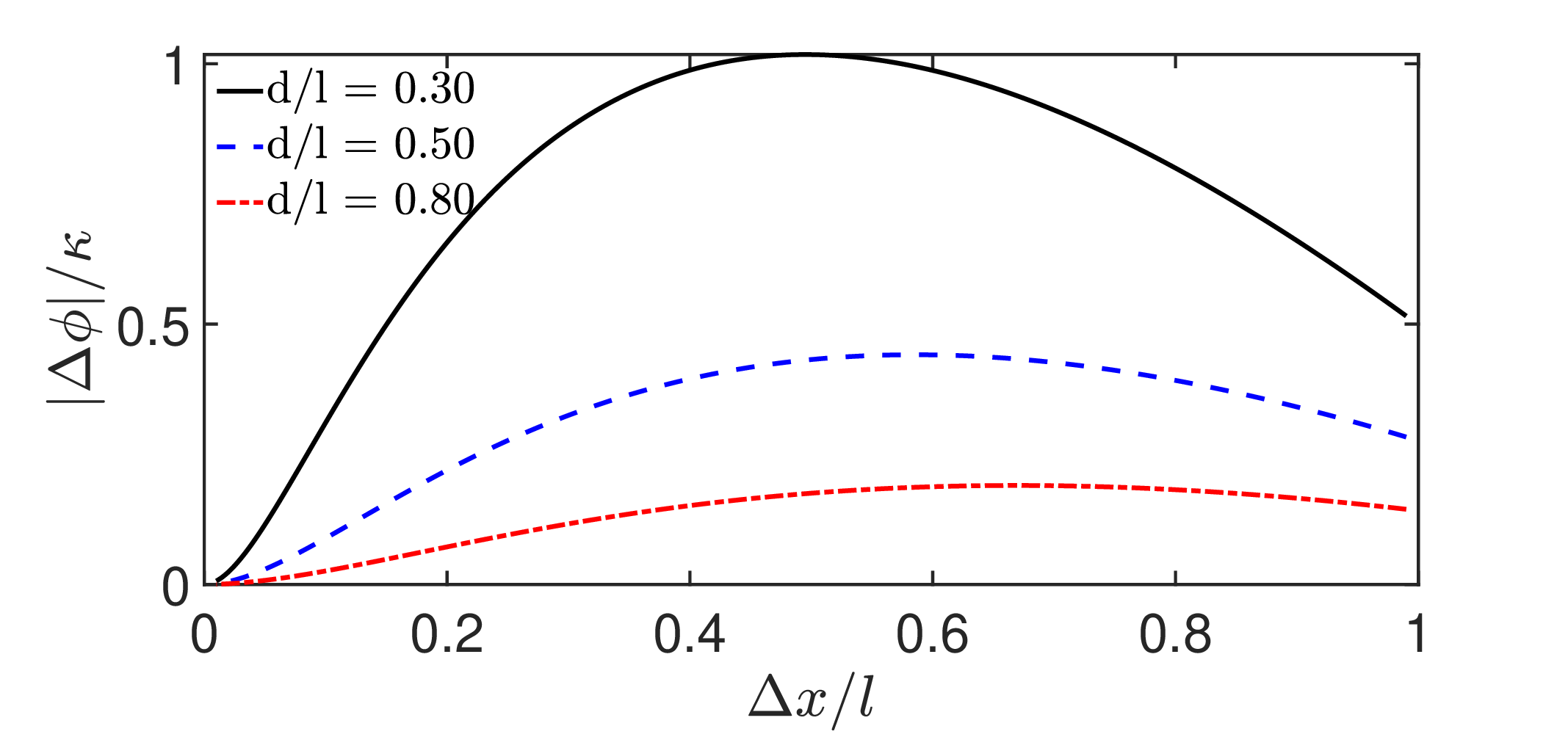}
	\caption{Rescaled induced phase $|\Delta\phi|/\kappa$ as a function of $\Delta x/l$ for several values of the normalized separation $d/l=0.3,\,0.5,\,0.8$ at fixed $\theta=\pi/2$.}
	\label{dBose}
\end{figure}

Figure~\ref{dBose} shows the dependence of the rescaled phase on the geometric parameter $\Delta x/l$ for different values of $d/l$. The phase exhibits a nonmonotonic behavior: it increases with $\Delta x/l$, reaches a maximum at an intermediate value, and then decreases. This indicates the presence of an optimal value of $\Delta x/l$ for each choice of $d/l$. The overall magnitude of the phase decreases with increasing $d/l$, consistent with the weakening of the gravitational interaction at larger separations.

\begin{figure}[t]
	\centering
	\includegraphics[width=0.45\textwidth]{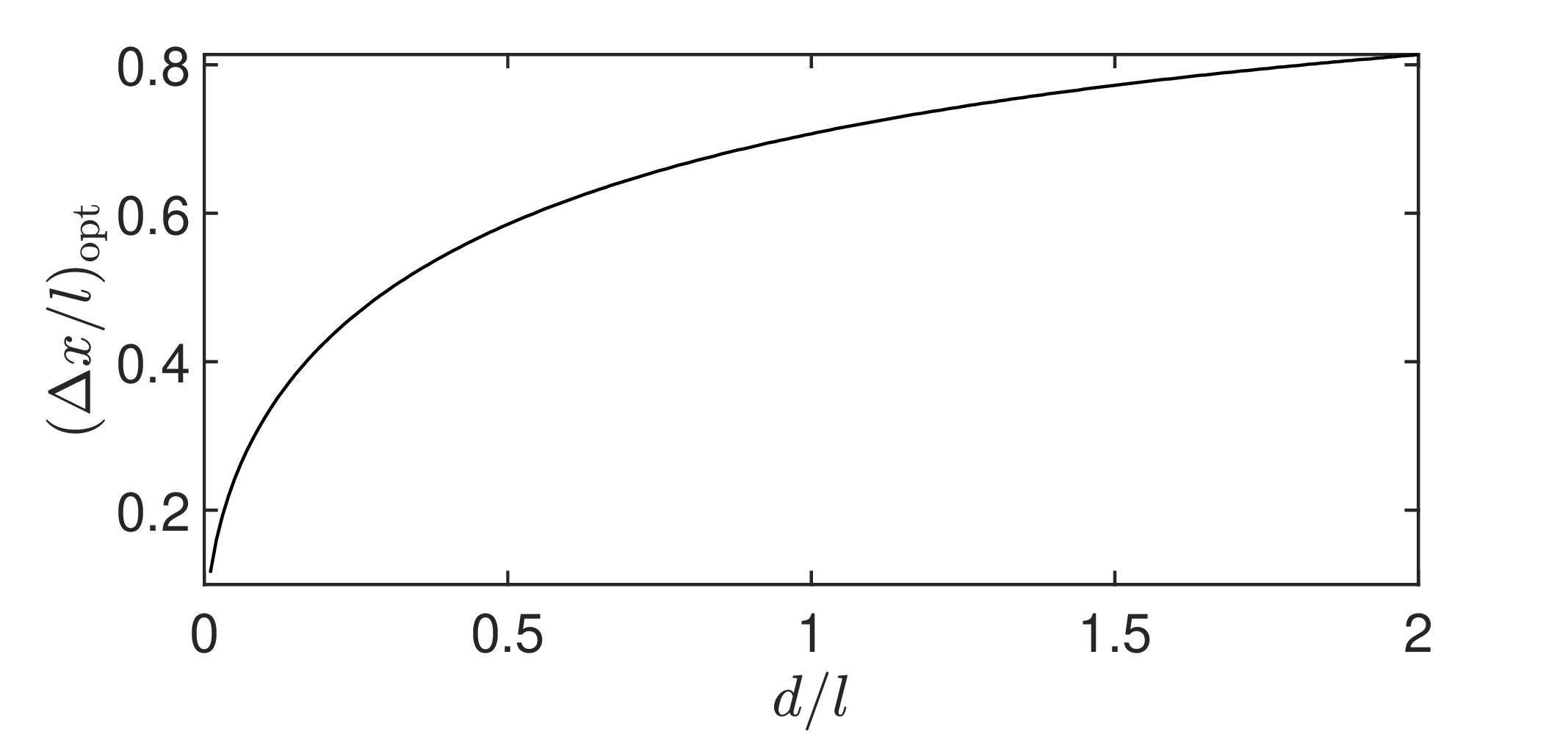}
	\caption{Optimal value $(\Delta x/l)_{\mathrm{opt}}$ that maximizes $|\Delta\phi|/\kappa$ as a function of the normalized separation $d/l$ for $\theta=\pi/2$.}
	\label{AoptBose}
\end{figure}

Figure~\ref{AoptBose} shows the value of $\Delta x/l$ that maximizes the rescaled phase as a function of $d/l$. The optimal parameter $(\Delta x/l)_{\mathrm{opt}}$ increases monotonically with increasing $d/l$. In practice, however, the minimum achievable separation is limited by non-gravitational forces, such as van der Waals interactions, which dominate at short distances \cite{schut2023relaxation}. As a result, there is an experimental cutoff that prevents the particles from being brought arbitrarily close to each other \cite{schut2023relaxation,bose2509spin}, and the optimal value of $\Delta x/l$ must be chosen according to the accessible range of $d/l$.

Overall, the induced phase—and therefore the generated entanglement—is enhanced for configurations with $\theta \approx \pi/2$ and appropriately chosen $\Delta x/l$, subject to the experimental constraints on the minimum separation.

 \section{The Second Experimental Setup}

\begin{figure}[h]
	\centering
	\includegraphics[width=3.5in]{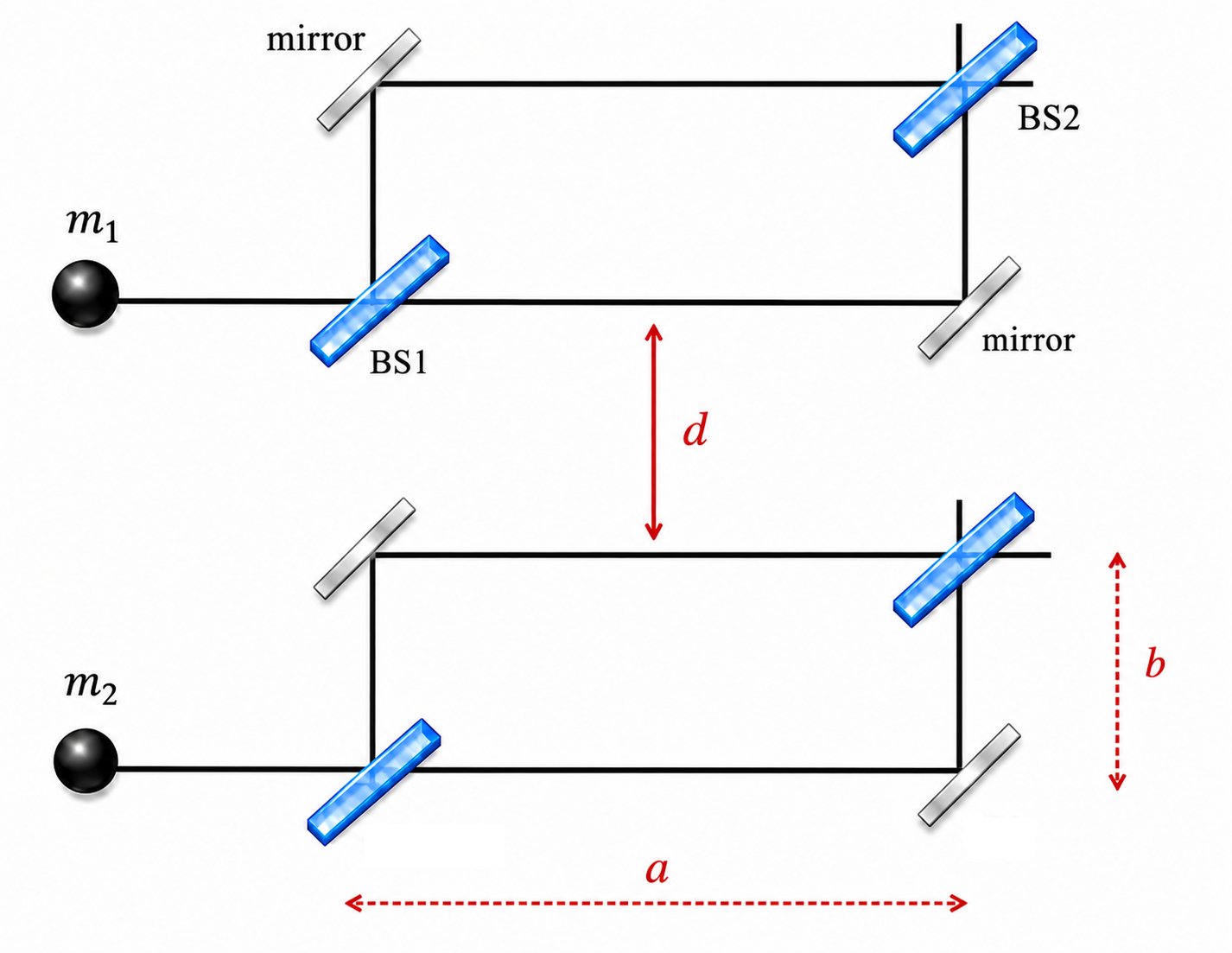}
	\caption{Experimental setup for generating gravitationally induced quantum entanglement using two Mach-Zehnder interferometers for massive particles.}
	\label{Fig2}
\end{figure}

For the next setup, we consider two massive particles, each passing through a Mach-Zehnder interferometer, as depicted in Fig.~\ref{Fig2}, which has been explored in recent literature \cite{marletto2017gravitationally,weber2024bose,di2024bose}. In this configuration, the horizontal arm of the interferometer has a length $a$, and the vertical arm has a length $b$. The total path length traversed by a particle before reaching the detector is given by $l = a + b$. The two identical interferometers are separated by a fixed distance $d$. 

Similar to the Stern-Gerlach setup, the upper and lower paths in the interferometer correspond to the orthogonal states $\vert u_i\rangle$ and $\vert d_i\rangle$ for the $i$-th interferometer ($i=1,2$). The state of the system just before the particles enter the second beam splitters in their respective interferometers evolves as given in Eq.\ref{3ifnt}. 
In this interferometer, it is also typically assumed that the lengths of the interferometers are much greater than their widths. Consequently, the gravitational interaction along the horizontal paths is considered the dominant contribution to phase accumulation. Under these conditions, the accumulated phases are expressed as 

\begin{align}
	\nonumber
	\phi_1 &= \phi_4 =  G\frac{m_1m_2T}{\hbar (d+b)}, 
    \\  	\nonumber
	\phi_2 &=G\frac{m_1m_2T}{\hbar d}, 
    \\
	\phi_3 &=  G\frac{m_1m_2T}{\hbar (d+2b)}.
\end{align}

In this setup, as in the previous one, this approximation neglects the phase contributions from the vertical segments of the paths. However, we show that the vertical segments may contribute significantly to the phase accumulation, making them no longer negligible. By carefully adjusting the configuration, an optimal setup for maximizing entanglement can be achieved. Again, we assume that the particles travel at a constant speed throughout their paths.

To this end, we consider the exact solution to the problem, incorporating contributions from the vertical paths. Taking these factors into account, the gravitationally induced phases $\phi_1$ to $\phi_4$ are determined to be

\begin{align}
	\phi_1 &= \phi_4 = G\frac{m_1m_2}{\hbar v} \frac {l}{d+b}\\
	\phi_2 &= G\frac{m_1m_2}{\hbar v} \times\\ 	\nonumber
	&\begin{cases}
		2g(b) + \frac{l-2b}{\sqrt{d^2+b^2}} , & \text{if $b \leq l/2$}, \\
		2g(l-b) +\frac{2b-l}{\sqrt{(d+2b-l)^2+(l-b)^2}}, & \text{otherwise},
	\end{cases} \\
\phi_3&= G\frac{m_1m_2}{\hbar v} \times\\ 	\nonumber
&\begin{cases}
	2f(b,d+b) + \frac{l-2b}{\sqrt{(d+2b)^2+b^2}} , & \text{if $b \leq l/2$}, \\
	2f(l-b,d+b) +\frac{2b-l}{\sqrt{(d+l)^2+(l-b)^2}}, & \text{otherwise},
\end{cases}
\end{align}

where
\begin{align}
	\nonumber
	f(x, y) &= \frac{1}{\sqrt{2}} \ln \left( \frac{2x + y + \sqrt{2(x + y)^2 + 2x^2}}{y(1 + \sqrt{2})} \right), \\ 
	g(x)&= \frac{1}{\sqrt{2}} \left( \sinh^{-1}\left(\frac{2x}{d + b} - 1\right) + \sinh^{-1}\left( 1\right) \right)
\end{align}
Now, we illustrate how optimizing the geometric configuration of the interferometer setup can significantly enhance the gravitationally induced phase, thereby maximizing the quantum entanglement. Conversely, suboptimal geometric configurations may reduce or even nullify the induced entanglement. For this analysis, we assume that the total path traversed by the particles, comprising both the vertical and horizontal arms, is fixed, as shown in Fig.~\ref{Fig2}. For a fixed total path length, we investigate the optimal design of the interferometer to achieve maximum entanglement.

The key aspect of this optimization lies in adjusting the vertical arm length, $b$, relative to the total length, $l$. Commonly, it is assumed that $l \gg b$, allowing the contribution of the vertical segment to be neglected. However, for optimal design, the critical parameter to consider is the ratio $b/l$.

\begin{figure}[t]
	\centering
	\includegraphics[width=3 in]{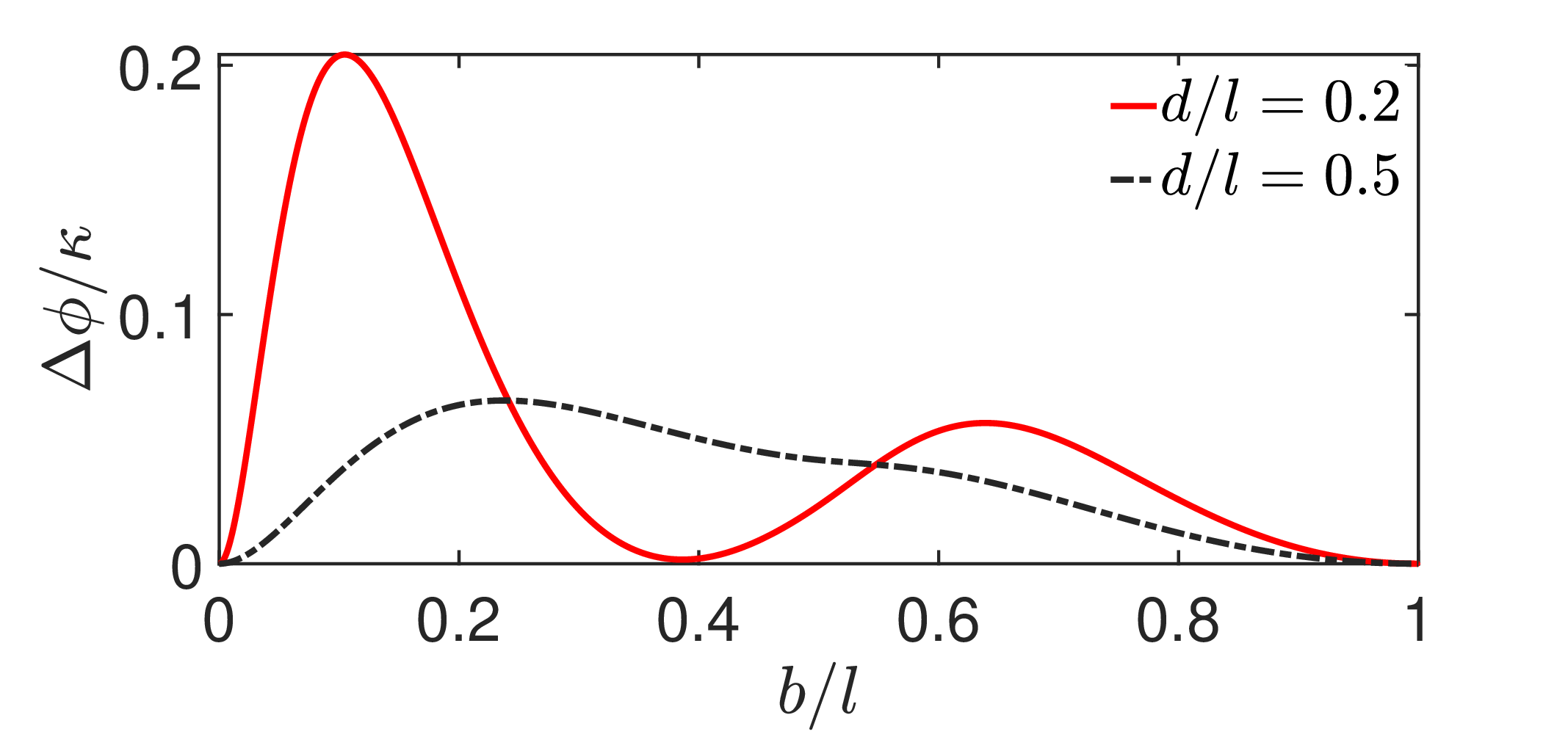}
	\caption{The rescaled induced phase $\Delta\phi/\kappa$ as a function of the relative vertical arm length $b/l$. The distance between the interferometers is represented by $d/l$, with $d/l = 0.5$ (dash-dot line) and $d/l = 0.2$ (solid line).
	}
	\label{dVedral}
\end{figure}

In Fig.~\ref{dVedral}, we plot the rescaled induced phase $\Delta\phi/\kappa$ as a function of $b/l$. For $d/l = 0.2$, the optimal configuration is achieved at $b/l = 0.10$, leading to a rescaled phase accumulation of approximately $0.20$. Conversely, at $b/l = 0.38$, the induced phase  becomes effectively negligible. These results highlight the importance of geometric optimization, as neglecting these considerations could lead to null results for certain parameter ranges. Similarly, for $d/l = 0.5$, the optimal phase occurs at $b/l = 0.23$, corresponding to a rescaled value of $0.07$. This demonstrates that fine-tuning the vertical arm length within a specific range can greatly enhance the induced phase.

Moreover, the results indicate that the optimal vertical arm length depends on the separation between the interferometers. Deviations from the optimal $b/l$ ratio can significantly impact phase accumulation, underscoring the experimental challenges associated with very small $b/l$ values. Additionally, increasing the separation $d$ between the interferometers reduces the induced phase, which aligns with the expectation that gravitational interactions weaken with increasing distance.

\begin{figure}[t]
	\centering
	\includegraphics[width=3 in]{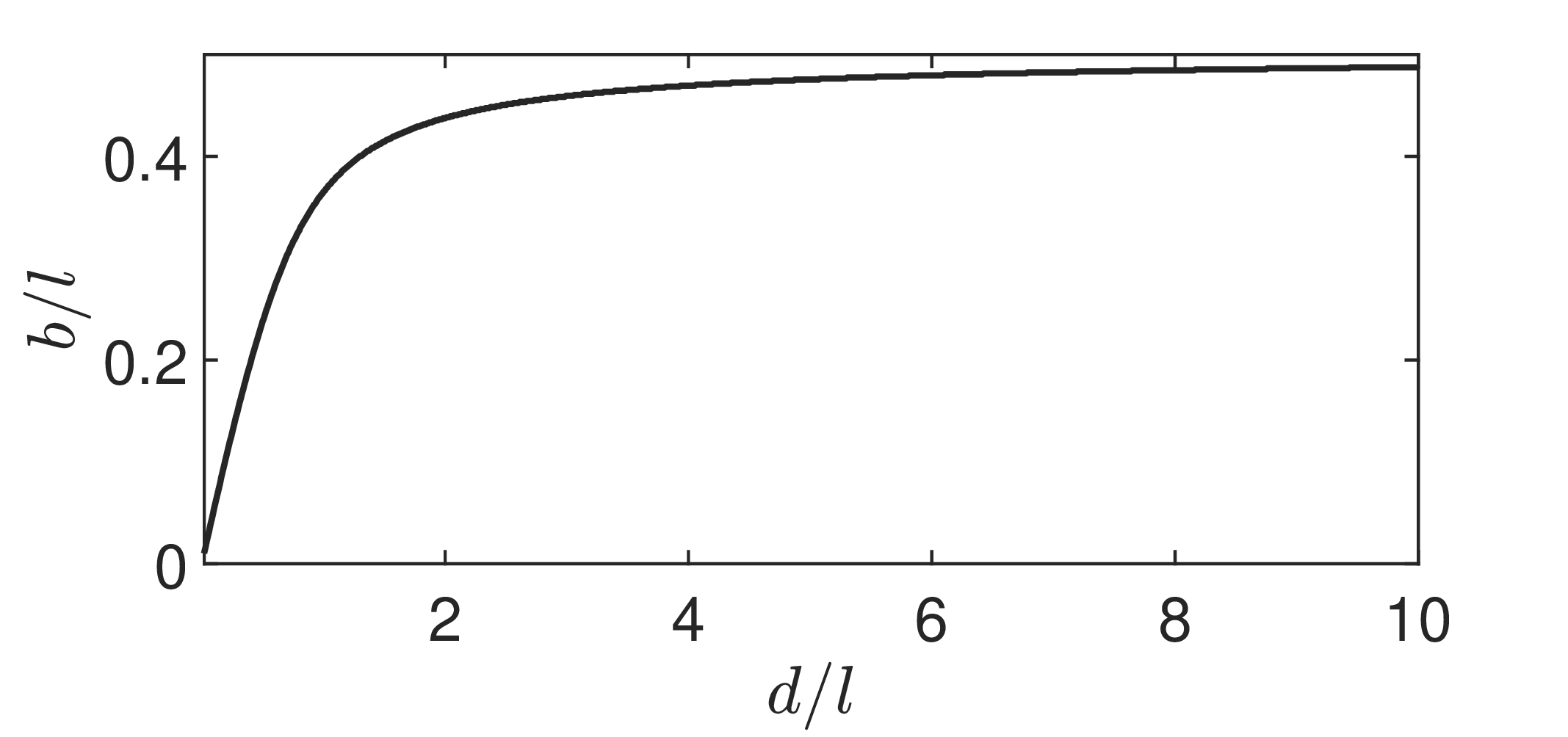}
	\caption{Optimal vertical arm length $b/l$ as a function of the relative separation between the interferometers, $d/l$. Both lengths are normalized to the total path length $l$.
	}
	\label{optVedral}
\end{figure}

To further explore the relationship between the interferometer separation and optimal geometry, Fig.~\ref{optVedral} shows the optimal $b/l$ values as a function of $d/l$. The analysis reveals that smaller $d/l$ ratios correspond to smaller optimal $b/l$ values. For much larger $d/l$, the optimal ratio gets closer to $b/l = 0.5$, which is a square configuration. These results emphasize the significance of the vertical segment in determining the total phase.

To ensure significant phase accumulation in the Mach-Zehnder interferometer, the experimental design should minimize the separation $d$ between the interferometers (see Fig.~\ref{dVedral}). Once the smallest feasible $d$ is selected, the geometry of the interferometers must be fine-tuned to the optimal $b/l$ ratio, as illustrated in Fig.~\ref{optVedral}, to maximize the gravitationally induced phase. \\

The analysis of the two interferometric configurations demonstrates that the geometric design of the setup plays a crucial role in determining the magnitude of gravitationally induced quantum entanglement. By incorporating the contributions from the vertical segments of the particle trajectories, we obtained exact expressions for the gravitational phases in both the Stern–Gerlach and Mach–Zehnder configurations. The results show that neglecting these contributions, as is commonly assumed in simplified treatments, can lead to a significant underestimation of the accumulated phase. Our analysis reveals the existence of optimal geometric parameters that maximize the phase difference, and therefore the attainable entanglement, for a fixed total path length. These findings highlight that careful geometric optimization of interferometric experiments can substantially enhance the observability of gravitationally induced entanglement. \\

\section{Summary and Conclusion}
\label{Conclusion}

This study investigated the gravitationally induced quantum entanglement in systems involving massive particles. By carefully analyzing the phase accumulation induced by gravitational interactions, we obtained important insights into the design and optimization of experimental setups aimed at detecting quantum gravitational effects.

In this work, we revisited the theoretical foundations and experimental feasibility of observing gravitationally induced quantum entanglement within a two-mass interferometric configuration, by deriving exact analytical expressions for the induced phases. These phases, which arise from the gravitational coupling between the two masses, are essential for the generation of entanglement. Our analysis highlights that neglecting the contribution of non-horizontal arms in the interferometers introduces significant challenges that may undermine the main goal of the experiment by obstructing the entanglement generation. This finding underscores the importance of accounting for all elements of the system in order to achieve reliable and meaningful results. Importantly, our study demonstrates that optimizing the geometry of the setup can significantly enhance the induced phase and thereby improve the resulting entanglement. This geometric optimization offers a practical and efficient approach to increasing experimental sensitivity and precision.

Since the experimental observation of gravitationally induced entanglement remains exceptionally challenging, requiring the preparation of massive quantum systems in coherent superposition states, suppression of dominant electromagnetic backgrounds, and preservation of coherence over extended timescales, any strategy that enhances the gravitational signal without introducing additional experimental complexity is particularly valuable. In this context, geometric optimization provides a comparatively accessible route to improving experimental sensitivity. We therefore expect that the present analysis can serve as useful guidance for the design of future experiments aimed at probing the quantum nature of gravity.
\section*{Acknowledgments}
The author thanks Y. Maleki for insightful discussions and valuable comments. The author is also grateful to the anonymous reviewer for their constructive feedback during the review process.

\bibliographystyle{apsrev4-1}
\bibliography{library}

\end{document}